\documentclass[twoside,11pt]{article}

\usepackage[cp1251]{inputenc}

\usepackage{graphicx}

\usepackage{amssymb}
\usepackage{amsmath}
\usepackage{amsfonts}
\usepackage{wasysym}

\begin{document}           

\title{\Large 
Quantum geometrodynamical description of the dark sector of the matter-energy content of the universe
}
\author{V.E. Kuzmichev, V.V. Kuzmichev\\[0.5cm]
\itshape Bogolyubov Institute for Theoretical Physics,\\
\itshape National Academy of Sciences of Ukraine, Kiev, 03680 Ukraine}

\date{}

\maketitle

\begin{abstract}
The evolution of the universe is studied in exactly solvable dynamical quantum model with the Robertson-Walker metric.
It is shown that the equation of motion which describes the expansion or contraction of the universe can be represented in the form
of the law of conservation of zero total energy for a particle with arbitrary mass being an analogue of the universe.
The analogue particle moves in the potential well under the action of the internal force produced by the curvature of space,
matter, and pressures of classical and quantum gravitational sources.
At a definite stage of the evolution of the universe,
this force can perform the positive work on the universe, which is similar to the work of the repulsive forces 
of dark energy, or it does the negative work analogous to the work of the attractive forces of dark matter.
The cases of real and complex state vectors which describe the geometrical properties of the universe filled with dust and radiation
are considered. It is shown that predictions of the quantum model do not contradict the observational data about the accelerating
expansion of our universe.
\end{abstract}


\section{Introduction}
The application of general relativity to understand modern astrophysical data about the large-scale structure and properties
of our universe leads to the necessity to postulate the existence of two dark substances, namely dark matter, in order to explain
the structure formation, and dark energy to justify the observed faintness of the extra-galactic SNe Ia, which is considered as  
the evidence of accelerating expansion of the present-day universe \cite{Ben,Pla}.
Concerning the physical properties of these substances, it is known that dark matter is gravitating, while
dark energy is antigravitating. The competition between these two components of the dark sector of matter-energy in the universe
determines the dynamics of the expansion which can change with time from deceleration, when dark matter is dominated, to 
acceleration, when the repulsive action of dark energy is predominant. Theoretically, one cannot exclude the inverse effect, when 
acceleration changes with time to deceleration, and the presence of the epoch, in which the corresponding forces are put in equilibrium. 

Since the physical nature of dark matter and dark energy remains unknown for now, numerous different models were proposed
in terms of the field-theoretical concept (see, e.g., Refs. \cite{Pee,Ber}). These models are intended to reconcile the classical theory of gravity, 
based on general relativity, with current astrophysical data.

At the same time, it is conceivable that the transition from decelerating expansion to accelerating expansion occurred at redshift
$\approx 0.6$, as well as the transition from inflation to radiation-domination, are not caused by the action of competing hypotetical
substances. It could be a reflection of the internal property of the universe which demonstrates that the universe is a more
complicated system than it is supposed in general relativity. For example, the universe could be a quantum object.

The aim of the present paper is to explain, on the basis of strict equations of quantum geometrodynamics for the specific, rather simple,
exactly solvable cosmological model, the possible change of regime of the expansion of the universe without the introduction ad hoc of such 
substances as dark matter and dark energy. We show that the dark sector of matter-energy may point to an existence of a particular type of 
forces acting in the universe. It is indicated that the nature of these forces is quantum.

In Sect.~2, we shortly review the Hamiltonian formalism for the minisuperspace model based on the Dirac-Arnowitt-Deser-Misner approach
to general relativity \cite{Dir58,ADM}, expounded in Refs. \cite{Kuz08,Kuz13}. The canonical quantization of matter and gravitational fields is 
given in Sect.~3.
Here, we demonstrate that, finally, the problem of the dynamics of the universe can be reduced to the problem of one-dimensional motion
of the analogue particle with arbitrary mass and zero total energy in the force field of the potential formed by the curvature of space,
matter, and quantum additions to the energy density and pressure of matter which are calculated precisely.
In Sect.~4, the specific quantum model of the universe with matter in the form of dust is studied.
The cases of real (universe of type I) and complex (universe of type II) state vectors are considered. The results obtained in the paper 
are summarized in Sect.~5.

Throughout the paper, the Planck system of units is used. As a result, all quantities in the equations become dimensionless.
The length $l_{P} = \sqrt{2 G \hbar / (3 \pi c^{3})}$ is taken as a unit of length and 
the $\rho_{P} = 3 c^{4} / (8 \pi G l_{P}^{2})$ is used as a unit of energy density and pressure. The proper time $\tau$ is
measured in units of length. An arc time (conformal time) $T$ is expressed in radians. 
The scalar field is taken in $\phi_{P} = \sqrt{3 c^{4} / (8 \pi G)}$, and so on. Here $G$ is Newton's gravitational constant.

\section{Hamiltonian Formalism}
In the present paper, we confine ourselves to a study of isotropic cosmological model. The space-time is described by the 
Robertson-Walker metric
\begin{equation}\label{1}
     ds^{2} = a^{2} [dT^{2} - d\Omega_{3}^{2}],
\end{equation}
where $a$ is the cosmic scale factor which is a function of time, $T$ is the time variable connected with the proper time $\tau$ 
by the differential equation $d\tau = a dT$, $T$ is the ``arc-parameter measure of time'': during the interval $d\tau$, a photon
moving on a hypersphere of radius $a(\tau)$ covers an arc $dT$ measured in radians \cite{MTW}.
$d\Omega_{3}^{2}$ is a line element on a unit three-sphere. Following the ADM formalism \cite{Dir58,ADM},
one can extract the so-called lapse function $N$, that specifies the time reference scale, from the total differential $dT$: $dT = N d \eta$,
where $\eta$ is the ``arc time'' which coincides with $T$ for $N = 1$ (cf. Refs. \cite{MTW,Lan2}). In the general case, the function $N$
plays the role of the Lagrange multiplier in the Hamiltonian formalism and it should be taken into account in an appropriate way.

To be specific, we consider the cosmological system (universe) described by the Hamiltonian \cite{Kuz08,Kuz13}
\begin{eqnarray}\label{2}
    H & = & \frac{N}{2} \left\{-\,\pi_{a}^{2} - a^{2} + a^{4} [\rho_{\phi} + \rho_{\gamma}]\right\} \nonumber \\ 
& + & \lambda_{1}\left\{\pi_{\Theta} - \frac{1}{2}\,a^{3} \rho_{0} s\right\}
+ \lambda_{2}\left\{\pi_{\tilde{\lambda}} + \frac{1}{2}\,a^{3} \rho_{0} \right\},
\end{eqnarray}
where $\pi_{a},\, \pi_{\Theta},\, \pi_{\tilde{\lambda}}$ are the momenta 
canonically conjugate with the variables $a,\, \Theta,\, \tilde{\lambda}$, $\rho_{\phi}$ is the energy density of matter (the field $\phi$),
$\rho_{\gamma}$ is the energy density of a perfect fluid, which defines a material reference frame \cite{Kuz08,Bro},
and it is a function of the density of the rest mass $\rho_{0}$ and the specific entropy $s$. 
$\Theta$ is the thermasy. $\tilde{\lambda}$ is the potential for the specific free energy taken with an inverse sign 
(for details see Ref. \cite{Kuz08}). The $N$, $\lambda_{1}$, and $\lambda_{2}$ are the Lagrange multipliers.

The Hamiltonian (\ref{2}) is a linear combination of constraints (expressions in braces) and thus weakly vanishes, $H \approx 0$.
The variations of the Hamiltonian with respect to $N$, $\lambda_{1}$, and $\lambda_{2}$ give three constraint equations,
\begin{equation}\label{2a}
-\,\pi_{a}^{2} - a^{2} + a^{4} [\rho_{\phi} + \rho_{\gamma}] \approx 0, \quad \pi_{\Theta} - \frac{1}{2}\,a^{3} \rho_{0} s \approx 0, 
\quad \pi_{\tilde{\lambda}} + \frac{1}{2}\,a^{3} \rho_{0} \approx 0.
\end{equation}
From the conservation of these constraints in time, it follows that the number of particles of a perfect fluid in the proper 
volume\footnote{This volume is equal to $2 \pi^{2} a^{3}$, where $a$ is taken in units of length.}
$\frac{1}{2} a^{3}$ and the specific entropy conserve: $E_{0} \equiv \frac{1}{2} a^{3} \rho_{0} = \mbox{const}$, $s =  \mbox{const}$.
Taking into account these conservation laws and vanishing of the momenta conjugate with the variables $\rho_{0}$ and $s$,
one can discard degrees of freedom corresponding to these variables, and convert the second-class constraints into
first-class constraints in accordance with Dirac's proposal \cite{Kuz08,Dir64}.

It is convenient to choose the perfect fluid with the density $\rho_{\gamma}$ in the form of relativistic matter (radiation).
Then, in Eq. (\ref{2a}) one can put $a^{4} \rho_{\gamma} \equiv E = \mbox{const}$. The matter field with the energy density 
$\rho_{\phi}$ and pressure $p_{\phi}$ can be taken for definiteness in the form of a uniform scalar field $\phi$,
\begin{equation}\label{3}
    \rho_{\phi} = \frac{2}{a^{6}}\,\pi_{\phi}^{2} + V(\phi), \quad p_{\phi} = \frac{2}{a^{6}}\,\pi_{\phi}^{2} - V(\phi),
\end{equation}
where $V(\phi)$ is the potential of this field, $\pi_{\phi}$ is the momentum conjugate with $\phi$.
After averaging with respect to appropriate quantum states, the scalar field turns into the effective matter fluid 
(see Ref. \cite{Kuz13}, and below).

The equation of motion for the classical dynamical variable $\mathcal{O} = 
\mathcal{O}(a, \phi, \pi_{a}, \pi_{\phi}, \dots )$ has the form
\begin{equation}\label{4}
    \frac{d\mathcal{O}}{dT} \approx \{\mathcal{O}, \frac{1}{N} H\},
\end{equation}
where $H$ is the Hamiltonian (\ref{2}), $\{.,.\}$ are the Poisson brackets. 

\section{Quantization}
In quantum theory, first-class constraint equations (\ref{2a}) become constraints on the state vector $\Psi$ \cite{Dir64}
and, in this way, define the space of physical states, which can be turned into a Hilbert space (cf. \cite{Kuc91}).
Passing from classical variables in Eqs. (\ref{2})-(\ref{3}) to corresponding operators, using the conservation laws, and
introducing the non-coordinate co-frame
\begin{equation}\label{4a}
    h d\tau = s d\Theta - d\tilde{\lambda}, \quad h dy = s d\Theta + d\tilde{\lambda},
\end{equation}
where $h = \frac{\rho_{\gamma} + p_{\gamma}}{\rho_{0}}$ is the specific enthalpy, $p_{\gamma}$ is the pressure of radiation,
and $y$ is a supplementary variable, we obtain \cite{Kuz08,Kuz13}
\begin{equation}\label{4b}
  \left( - \partial_{a}^{2} + a^{2} - 2 a \hat{H}_{\phi} - E \right) | \Psi \rangle = 0, \quad \partial_{y}| \Psi \rangle = 0,
  \quad \left(-i \partial_{T} - \frac{2}{3}E \right) | \Psi \rangle = 0,
\end{equation}
where
\begin{equation}\label{6a}
\hat{H}_{\phi} = \frac{1}{2} a^{3} \hat{\rho}_{\phi}
\end{equation}
is the operator of Hamiltonian of the scalar field $\phi$, the operator $\hat{\rho}_{\phi}$ is described by Eq. (\ref{3}) 
with $\pi_{\phi} = -i \partial_{\phi}$.
From Eq. (\ref{4b}), it follows that 
the evolution in time of the state vector $\Psi$ is described by the exponential multiplier as follows
\begin{equation}\label{5}
    \Psi (T) = \mbox{e}^{\,i\frac{2}{3}\,E (T - T_{0})} \Psi (T_{0}),
\end{equation}
so that arc-parameter $T$ appears to be the most natural time variable in quantum theory as well.
Here $T_{0}$ is an arbitrary constant taken as a time reference point. 
The vector $\Psi (T_{0}) \equiv |\psi\rangle$ 
is defined in the space of two variables  $a$ and $\phi$. According to Eqs. (\ref{4b}), it is annihilated by the constraint equation 
\begin{equation}\label{6}
\left( - \partial_{a}^{2} + a^{2} - 2 a \hat{H}_{\phi} - E \right) | \psi \rangle = 0.
\end{equation}

By substituting the Poisson brackets with commutators of operators $\hat{\mathcal{O}} = \{a, -i \partial _{a}\}$ and $\frac{1}{N} \hat{H}$, we obtain
the quantum analog of Eq. (\ref{4}) for the operator of momentum $\pi_{a}= -i \partial _{a}$ and its time derivative  
\begin{equation}\label{7}
\langle \psi |- i  \partial_{a}| \psi \rangle = \langle \psi |- \frac{da}{dT}| \psi \rangle,
\end{equation}
\begin{equation}\label{8}
\langle \psi |- i  \frac{d}{dT} \partial_{a}| \psi \rangle = \langle \psi |a - \hat{H}_{\phi} + 3 \hat{L}_{\phi} | \psi \rangle,
\end{equation}
where
\begin{equation}\label{9}
\hat{L}_{\phi} = \frac{1}{2} a^{3} \hat{p}_{\phi}
\end{equation}
is the operator of Lagrangian of the scalar field, and $\hat{p}_{\phi}$ is given by Eq. (\ref{3}) with $\pi_{\phi} = -i \partial_{\phi}$. 

The operator in the left-hand side of Eq.  (\ref{6}) is not the Hamiltonian of the system (it has the dimensions of [energy]$\times$[length] in
physical units). Whether this operator is self-adjoint depends on the behaviour of the vector $| \psi \rangle$ and its first derivatives with
respect to the scale factor and field variables on the boundaries of range of their values. In this connection, we consider
the Hamiltonian $\hat{H}_{\phi}$, which can be diagonalized by means of some state vectors $\langle x|u_{k} \rangle$ of quantum scalar field
in the representation of generalized variable $x = x(\frac{1}{2} a^{3}, \phi)$. The explicit form of $x$ is determined by the form of the potential 
$V(\phi)$ taken as a real function \cite{Kuz13}. 
Assuming that vectors $|u_{k} \rangle$ satisfy the completeness condition, $\sum_{k} |u_{k} \rangle\langle u_{k} | = 1$,
and that they are orthonormalized,  $\langle u_{k} |u_{k'} \rangle= \delta_{k k'}$, we guarantee the self-adjointness of the operator
$\hat{H}_{\phi}$ and reality of the function $M_{k} (a)$ in the equation
\begin{equation}\label{10}
\langle u_{k}| \hat{H}_{\phi} |u_{k'} \rangle = M_{k} (a) \delta_{k k'},
\end{equation}
where the index of the state $k$ can take both discrete and continuous values (in the latter case, the condition of orthogonality
of the state vectors $|u_{k} \rangle$ is written by means of the Dirac delta function),
$M_{k} (a) = \frac{1}{2} a^{3} \rho_{m}$ is the proper energy of matter in the volume $\frac{1}{2} a^{3}$. The energy density and pressure of matter\footnote{The index $k$ is omitted.},
\begin{equation}\label{11}
\rho_{m} = \langle u_{k} | \hat{\rho}_{\phi}| u_{k} \rangle, \quad
p_{m} = \langle u_{k} | \hat{p}_{\phi}| u_{k} \rangle,
\end{equation}
have the form
\begin{equation}\label{12}
\rho_{m} = \frac{2 M_{k}(a)}{a^{3}}, \quad
p_{m} = - \frac{2}{3 a^{2}} \frac{d M_{k}(a)}{da}.
\end{equation}
In the general case, the proper energy $M_{k} (a)$ depends on $a$. It describes a classical source (as a mass-energy) 
of the gravitational field in $k$th state.

Using Eq. (\ref{10}), one can integrate Eqs. (\ref{6})-(\ref{8}) with respect to the matter field variable.
Expressing the vector $| \psi \rangle$ in the form of expansion in terms of the complete set of states $| u_{k} \rangle$,
\begin{equation}\label{13}
|\psi \rangle = \sum_{k} |u_{k} \rangle \langle u_{k}  |\psi \rangle,
\end{equation}
from Eq. (\ref{6}), we obtain the equation for the function $\langle a | f_{k} \rangle \equiv \langle u_{k}  |\psi \rangle$,
\begin{equation}\label{14}
\left( - \partial_{a}^{2} + a^{2} - 2 a M_{k} (a) \right) | f_{k} \rangle = E | f_{k} \rangle.
\end{equation}
Since in this approach E is a real constant, the operator in the left-hand side of Eq. (\ref{14}) has to be self-adjoint.
Defining the flux with respect to the variable $a$,
\begin{equation}\label{14a}
J_{k}(a) = \frac{1}{2i} \left[\langle a | f_{k} \rangle^{*} \partial_{a} \langle a | f_{k} \rangle - 
\langle a | f_{k} \rangle \partial_{a} \langle a | f_{k} \rangle^{*} \right],
\end{equation}
we find that the condition of self-adjointness is satisfied, if $J_{k}(0) = J_{k}(\infty) = 0$. In this case, Eq. (\ref{14}) can be considered
as an eigenvalue equation. Its solution $| f_{k} \rangle$ is an eigenfunction corresponding to the eigenvalue $E$.
The function $| f_{k} \rangle$ describes the geometrical properties of the quantum universe filled with matter, whose mass-energy
is $M_{k}(a)$.

In order to turn to the classical observables (such as the Hubble expansion rate and deceleration parameter), 
we extract the amplitude and the phase $S_{k}(a)$ in the function $| f_{k} \rangle$
\begin{equation}\label{15}
\langle a | f_{k} \rangle = \frac{C_{k}}{\sqrt{\partial_{a} S_{k} (a)}}\ e^{i S_{k}(a)},
\end{equation}
where $C_{k}$ is the constant determined by the boundary condition on the function $\langle a | f_{k} \rangle$, e.g. on the asymptotics
$a \rightarrow \infty$. If the function $| f_{k} \rangle$ is real, then it is expressed through the Euclidean phase $S_{E} = - i S_{k}$.
If the phase $S_{k}$ is a real function, then Eq. (\ref{15}) will describe the outgoing or incoming wave propagating in the space
of the scale factor $a$. The general solution of Eq. (\ref{14}) in this case will be a superposition of $| f_{k} \rangle$- 
and $\langle f_{k} |$-states separately describing expanding or contracting quantum universe (cf. Ref. \cite{Har83}).

Substituting the expression (\ref{15}) into Eq. (\ref{14}) and taking into account that $\langle a | f_{k} \rangle$ is nontrivial, we obtain
the non-linear equation for the phase $S_{k}(a)$
\begin{equation}\label{22}
(\partial_{a} S_{k})^{2} + a^{2} - 2 a M_{k}(a) - E - \frac{3}{4} \left(\frac{\partial_{a}^{2} S_{k}}{\partial_{a} S_{k}} \right)^{2} 
+ \frac{1}{2} \frac{\partial_{a}^{3} S_{k}}{\partial_{a} S_{k}} = 0.
\end{equation}
This equation contains only the derivatives of the phase, therefore it could be considered as an equation for $\partial_{a} S_{k}$.
Using Eqs. (\ref{13}) and (\ref{15}), from Eq. (\ref{7}), we get the equality
\begin{equation}\label{22a}
\langle f_{k} | \left(\partial_{a} S_{k} + \frac{i}{2} \frac{\partial_{a}^{2} S_{k}}{\partial_{a} S_{k}} + \frac{da}{dT}\right) | f_{k} \rangle = 0.
\end{equation}
If the phase $S_{k}$ is a pure imaginary function of $a$ and the momentum $\pi_{a} = - \frac{da}{dT}$ is real, 
then the operator in parentheses will be self-adjoint, when the expectation value 
$\langle f_{k} |\frac{da}{dT} | f_{k} \rangle = 0$. In the general case, $\langle f_{k} |\left(\frac{da}{dT}\right)^{2} | f_{k} \rangle \neq 0$.
If the phase $S_{k}$ is a real function, then the condition of self-adjointness of the mentioned operator takes the form:
$\langle f_{k} | \frac{\partial_{a}^{2} S_{k}}{\partial_{a} S_{k}} | f_{k} \rangle = 0$. But, in general, 
$\langle f_{k} | \left( \frac{\partial_{a}^{2} S_{k}}{\partial_{a} S_{k}}\right)^{2} | f_{k} \rangle \neq 0$. Since the function $|f_{k} \rangle$
is nontrivial, from Eq. (\ref{22a}), it follows 
\begin{equation}\label{22b}
\partial_{a} S_{k} + \frac{i}{2} \frac{\partial_{a}^{2} S_{k}}{\partial_{a} S_{k}} = - \frac{da}{dT}.
\end{equation}
Using this relation, one can reduce Eq. (\ref{22}) to the form
\begin{equation}\label{16}
\frac{1}{2} \left(\frac{da}{dT}\right)^{2} + U(a) = 0,
\end{equation}
where
\begin{equation}\label{17}
U(a) = \frac{1}{2}\left[a^{2} - 2 a M_{k}(a) - Q_{k}(a) - E \right].
\end{equation}
The function 
\begin{equation}\label{18}
Q_{k}(a) = i \partial_{a}^{2} S_{k} + \frac{1}{2} \left[\left(\frac{\partial_{a}^{2} S_{k}}{\partial_{a} S_{k}} \right)^{2} 
- \frac{\partial_{a}^{3} S_{k}}{\partial_{a} S_{k}}\right]
\end{equation}
determines the quantum correction $\rho_{Q}$ to the energy density of matter in the form
\begin{equation}\label{19}
\rho_{Q} = \frac{Q_{k}(a)}{a^{4}} \equiv \frac{2 M_{Q}(a)}{a^{3}},
\end{equation}
where $M_{Q}(a) = \frac{1}{2} a^{3} \rho_{Q}$ is the proper energy of the quantum source of the gravitational field.
The pressure produced by the quantum source
\begin{equation}\label{20}
P_{Q} = - \frac{2}{3 a^{2}} \frac{d M_{Q}(a)}{da} \equiv p_{Q} + p_{Q\gamma}
\end{equation}
is the sum of the pressure $p_{Q}$, which is the quantum correction to the pressure $p_{m}$, and the correction for relativity $p_{Q\gamma}$,
\begin{equation}\label{21}
p_{Q} = - \frac{1}{3 a^{3}} \frac{d Q_{k}(a)}{da}, \quad p_{Q\gamma} = \frac{1}{3} \rho_{Q}.
\end{equation}

According to Eqs. (\ref{17})-(\ref{21}), all quantum corrections to the energy density and pressure of ordinary matter in the universe
are collected in the function of gravitational quantum source $Q_{k}(a)$.

Passing to dimensional physical units, we find \cite{Kuz13} that the first term in $Q_{k}$ is proportional to $l_{P}^{2}$, while
the term with higher derivatives of the phase $S_{k}$ in square brackets of Eq. (\ref{18}) is proportional to $l_{P}^{4}$.
Therefore, one can conclude that quantum corrections make contributions $\sim \hbar$ and $\sim \hbar^{2}$ to the dynamics
of the expanding universe. 

From Eq. (\ref{16}), after differentiation with respect to $T$, it follows
\begin{equation}\label{16a}
\frac{d^{2}a}{dT^{2}} = - \frac{dU(a)}{da}.
\end{equation}

The formulae (\ref{16}) and (\ref{16a}) allow to draw an analogy with the equations of classical mechanics describing the conservation of energy 
of a particle moving in the potential well (\ref{17}). These relations may be interpreted as the equations which describe the motion of the 
particle, an analogue of the universe, with an arbitrary mass and zero total energy under the action of the force
\begin{equation}\label{23}
F(a) = - \frac{dU(a)}{da} = -a + M_{k}(a) + a \frac{dM_{k}(a)}{da} + \frac{1}{2} \frac{dQ_{k}(a)}{da}.
\end{equation}
In addition to the effect of curvature of space and the mass term, this force takes into account the gradients (pressures) of
classical and quantum gravitational sources. 

It should be pointed out that the relations (\ref{16}) and (\ref{16a}) only formally coincide with the equations of classical mechanics. 
They describe the universe in which, in addition to classical source of gravitational filed in the form of matter with the mass
$M_{k}(a)$, there is relativistic quantum source with the mass $M_{Q}(a)$ which, under specific conditions, can have a serious influence on
the dynamics of the universe. These conditions depend on the relation between the masses $M_{k}(a)$ and $M_{Q}(a)$.
The mass $M_{k}(a)$ is given by the Hamiltonian $\hat{H}_{\phi}$ (\ref{6a}), i.e., in the end, by the potential $V(\phi)$ chosen from model 
arguments. The mass $M_{Q}(a)$ is defined by the function of quantum source $Q_{k}(a)$, whose form (\ref{18}) is totally determined by the 
solution of the quantum problem (\ref{22}).

Passing to the proper time $\tau$, one reduces the equations (\ref{16}) and (\ref{16a}) to
\begin{equation}\label{23a}
\left(\frac{\dot{a}}{a}\right)^{2} = \rho_{tot} - \frac{1}{a^{2}}, \quad \frac{\ddot{a}}{a} = - \frac{1}{2} \left(\rho_{tot} + 3 p_{tot} \right),
\end{equation}
where dots denote the derivatives with respect to $\tau$, and 
\begin{equation}\label{23b}
\rho_{tot} = \rho_{m} + \rho_{\gamma} + \rho_{Q}, \quad p_{tot} = p_{m} + p_{\gamma} + P_{Q}.
\end{equation}

The deceleration parameter $q = - \frac{a \ddot{a}}{\dot{a}^{2}}$ in the model under consideration is reduced to the expression
\begin{equation}\label{31}
q = 1 - \frac{a}{2 U} \frac{d U}{d a}.
\end{equation}

In the approximation $Q_{k} = 0$, the relations (\ref{23a}) and (\ref{23b}) reduce to the ordinary Einstein-Friedmann equation 
which describes the closed universe filled with matter with the density $\rho_{m}$ and radiation with the density
$\rho_{\gamma}$.

In order to clarify the physical reason of the origin of the quantum correction $p_{Q}$ (\ref{21}) to the total pressure in the universe,
let us consider the consequences which follow from Eqs. (\ref{7}) and (\ref{8}).

According to the well-known rule, the time derivative of the quantum mechanical mean value of a given operator is equal to 
the mean value of the time derivative of this operator \cite{Lan65}. Thus, taking into account Eq. (\ref{7}), we reduce Eq. (\ref{8})
to the form
\begin{equation}\label{24a}
\langle \psi |\left(- \frac{d^{2}a}{dT^{2}} - a + \hat{H}_{\phi} - 3 \hat{L}_{\phi}\right) | \psi \rangle = 0.
\end{equation}
Using the expression (\ref{13}), integrating with respect to $x$, substituting the expression (\ref{15}), and taking into account 
the self-adjointness of the operator in parentheses in Eq. (\ref{24a}) and the non-triviality of the function $\langle a| f_{k}\rangle$,
we obtain
\begin{equation}\label{24b}
\frac{d^{2}a}{dT^{2}} = - a + M_{k}(a) + a\frac{dM_{k}(a)}{da} - \frac{3}{2} a^{3} \bar{p}_{Q},
\end{equation}
where 
\begin{equation}\label{24c}
\bar{p}_{Q} = \sum_{k' \neq k} \langle u_{k}|\hat{p}_{Q}| u_{k'} \rangle \frac{C_{k'}}{C_{k}} 
\sqrt{\frac{\partial_{a} S_{k}}{\partial_{a} S_{k'}}} e^{i(S_{k'} - S_{k})}.
\end{equation}
Comparing this expression with Eqs. (\ref{16a}) and (\ref{23}), and using Eq. (\ref{21}), we find that
\begin{equation}\label{24d}
\bar{p}_{Q} = p_{Q}.
\end{equation}
It means that the quantum correction to the pressure of matter is stipulated by the fact that the state vector $|\psi \rangle$ (\ref{13})
is the superposition of all possible states of the classical source of the gravitational field $M_{k}(a)$. In our approach,
$p_{Q}$ is the only quantity which takes into account that the probability amplitudes (\ref{15}) mix.

The equations (\ref{16}) and (\ref{16a}) are exact. From these equations, it follows that, in general case, the force (\ref{23}) can
perform both the positive work on the universe, which is similar to the work of the repulsive forces of dark energy, and
the negative work analogous to the work of the attractive forces of dark matter. The kind of work which is performed on the universe
depends on the sign and behaviour of the potential well $U(a)$ in Eq. (\ref{16}).

In order to find out the impact of the mass $M_{k} (a)$ and function of quantum source $Q_{k}(a)$ on the evolution of the universe,
we consider a specific exactly solvable quantum problem.

\section{An exactly solvable model}
Let matter be represented by a dust ($p_{m} = 0$).
Such a type of matter is reproduced by the scalar field model with the potential $V(\phi) = \lambda \phi ^{2}$, 
where the field $\phi$ oscillates near the point of its true vacuum, $\lambda$ is the coupling constant \cite{Kuz13}.

Really, if one introduces the variable $x = \left(\frac{\lambda a^{6}}{2} \right)^{1/4} \phi$, then the Hamiltonian $\hat{H}_{\phi}$ (\ref{6a})
takes the form
\begin{equation}\label{31a}
\hat{H}_{\phi} = \left(\frac{\lambda}{2} \right)^{1/2} \left(-\partial_{x}^{2} + x^{2} \right).
\end{equation}
We introduce the state vectors $\langle x|u_{k}\rangle$, which satisfy the equation
\begin{equation}\label{31b}
\left(-\partial_{x}^{2} + x^{2} - \epsilon_{k} \right)|u_{k}\rangle = 0,
\end{equation}
where $\epsilon_{k}$ is an eigenvalue. This equation describes the quantum oscillator with $\epsilon_{k} = 2 k + 1$, $k = 0, 1, 2, \ldots$.
From Eqs. (\ref{10}), (\ref{31a}), and (\ref{31b}), it follows that 
\begin{equation}\label{32}
M_{k}(a) = \sqrt{2 \lambda} \left(k + \frac{1}{2} \right) \equiv M.
\end{equation}
Here $M$ is the total mass of $k$ non-interacting identical particles with the masses $\sqrt{2 \lambda}$.

It is convenient to introduce a new variable $z = a - M$ which describes a deviation of $a$ from its ``equilibrium'' value at the point,
where\footnote{In dimensional units, we have $a = \frac{2}{3 \pi} \frac{G}{c^{2}} M$ (cf. \cite{MTW,Lan2}).} $a = M$.
Then, Eqs. (\ref{14}) and (\ref{22}) take the form (the index $k$ is omitted)
\begin{equation}\label{33}
\left[ - \partial_{z}^{2} + z^{2} - (2 n + 1) \right] | f \rangle = 0,
\end{equation}
\begin{equation}\label{34}
(\partial_{z} S)^{2} + z^{2} - (2 n + 1) = \frac{3}{4} \left(\frac{\partial_{z}^{2} S}{\partial_{z} S} \right)^{2} 
- \frac{1}{2} \frac{\partial_{z}^{3} S}{\partial_{z} S},
\end{equation}
where $n = 0, 1, 2, \ldots$ is the quantum number which numerates the discrete states of the universe,
$E + M^{2} = 2 n + 1$, in the potential well $z^{2}$.

The potential well (\ref{17}) in Eqs. (\ref{16}) reduces to
\begin{equation}\label{35}
U = \frac{1}{2}\left[z^{2} - (2 n + 1) - Q(z) \right].
\end{equation}
The quantities $| f \rangle$, $S$, and $U$ are the functions of $z$. Also they depend on free indices $k$ and $n$ which are
omitted here and below, when these indices are inessential.

Both equations (\ref{33}) and (\ref{34}) have two solutions
\begin{equation}\label{36}
\langle z | f \rangle_{1} = H_{n}(z) e^{-z^{2}/2}, \quad
\langle i z | f \rangle_{2} = H_{-n-1}(iz) e^{z^{2}/2},
\end{equation}
and
\begin{equation}\label{37}
\partial_{z} S_{1}(z) = i \frac{e^{z^{2}} H_{n}^{-2}(z)}{2 \int_{0}^{z} dx\, e^{x^{2}} H_{n}^{-2}(x)},
\end{equation}
\begin{equation}\label{38}
\partial_{z} S_{2}(i z) = - \frac{e^{-z^{2}} H_{-n-1}^{-2}(i z)}{2 \int_{0}^{i z} dx\, e^{x^{2}} H_{-n-1}^{-2}(x)},
\end{equation}
respectively, where $H_{\nu}(y)$ is the Hermitian polynomial.
According to (\ref{15}) and (\ref{36}), the function $| f \rangle_{1}$ is real and expressed through the Euclidean phase
$S_{E} = - i\,S_{1}$. The second solution $| f \rangle_{2}$ of Eq. (\ref{33}) is complex. The corresponding phase $S_{2}$ appears to be complex.
Usually the solution $| f \rangle_{2}$ is discarded as unphysical. However, in quantum cosmology both solutions should be considered, since
only in such an approach, one can obtain nontrivial results about topological properties of the universe as an essentially quantum system and
clarify the nature of dark matter and dark energy.

\subsection{The universe of type I}
Let us consider the first solution $| f \rangle_{1}$ (universe of type I). Substituting $\partial_{z} S_{1}(z)$ into Eq. (\ref{18}), we obtain
the expression for the function of quantum source,
\begin{equation}\label{39}
Q(z) = - (2 n + 1) + 2 n\, \frac{H_{n-1}(z) H_{n+1}(z)}{H_{n}^{2}(z)}.
\end{equation}
Then, from Eq. (\ref{35}), we find the potential well as the function of the deviation $z$,
\begin{equation}\label{40}
U(z) = \frac{1}{2} z^{2} - n\, \frac{H_{n-1}(z) H_{n+1}(z)}{H_{n}^{2}(z)}.
\end{equation}
This energy depends on quantum $n$th state of the universe. This state is determined by the mass $M$ of the universe in accordance
with the condition of quantization: $2n + 1 = M^{2} + E$. For example, the observed part of our universe is characterized by the parameters
$M \sim 10^{61}$ ($\sim 10^{80}$ GeV) and $E \sim 10^{118}$ ($\rho_{\gamma} \approx 10^{-10}$ GeV/cm$^{3}$) \cite{MTW,PDG}.
From the viewpoint of the model under consideration, it is in the state with $n \sim 10^{122}$ (up to $\sim 10^{-4}$).
This estimate practically coincides with the estimate given by Hartle and Hawking \cite{Har83}.

The function (\ref{40}) for the specific value $n = 10$, chosen for the illustration of a general regularity, is shown in Fig.~1.
The main features of the behaviour of the potential well $U(z)$ and other parameters remain intact up to the values $n \gg 1$
acceptable for the present-day universe.
In this example, the mass of the universe equals to $M = 4.08$.
The function $U(z)$ is nonnegative for all possible values of the deviation $z$. Its range is partitioned by the points $z_{i}$, where 
$U(z_{i}) = 0$, into $n$ nonoverlapping consecutive intervals. The potential well $U (z)$ goes to infinity at the points where the function 
$| f \rangle_{1}$ vanishes. 

From the conservation law of zero total energy (\ref{16}), it follows that the kinetic energy $\frac{1}{2}(\frac{dz}{dT})^{2}$ is negative and
vanishes at the points, where the potential energy also vanishes. The negative kinetic energy of the analogue particle can be
interpreted as the positive energy of the particle moving in imaginary time $\xi = -i T$. The metric (\ref{1}) of such a universe
will have the Euclidean signature.

\begin{figure*}
\includegraphics[width=8cm]{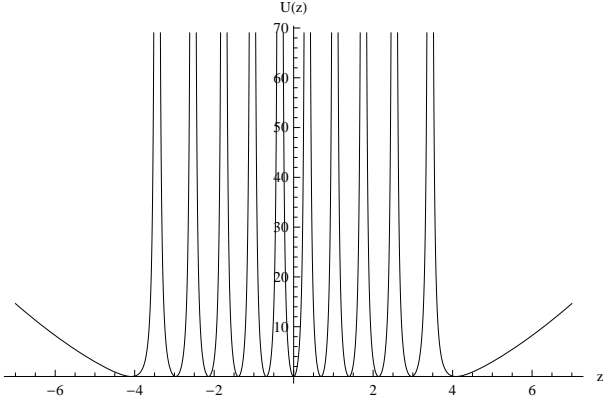}
\caption{The potential well $U(z)$ (\ref{40}) \textit{versus} the deviation $z$ for $n = 10$.}
\end{figure*} 

The formal solution of Eqs. (\ref{16}) for the potential well (\ref{40}) can be found in quadrature, 
\begin{equation}\label{41}
i T = \int\!dx \frac{H_{n}(x)}{\sqrt{x^{2} H_{n}^{2}(x) - 2n\,H_{n-1}(x) H_{n+1}(x)}} + const.
\end{equation}

The deceleration parameter (\ref{31}) for the potential well (\ref{40}) with $n = 10$ is plotted in Fig.~2. As we can see, 
the domains, where the expansion of the universe is accelerating or decelerating, follow each other sequentially, as $z$ increases.
They are separated by the points on the axis $z$, where $U(z)$ vanishes, while its derivative with respect to $z$
changes its sign. The deceleration parameter for the quantum universe in the vacuum state $n = 0$ has a simple form
$q_{n = 0} = - \frac{M}{z}$ and it demonstrates that near the point $z = - M$ (i.e. $a = 0$) the expansion is decelerating, $q = 1$.
At the point $z = 0$ ($a = M$) the parameter $q$ undergoes a jump from $q = + \infty$ to $q = - \infty$. For $z > 0$, the universe
expands with acceleration, which weakens with the increase of $z$ and is characterized by
the parameter $q = - 1$ on the boundary $z = M$. For the highly excited states, the early universe (small $a$, not far from $z = - M$)
can be in the state, where the universe accelerates or decelerates, while in the late universe (at $a \gtrsim 2 M$ or $z \gtrsim M$), the
expansion is accelerating.

\begin{figure*}
\includegraphics[width=8cm]{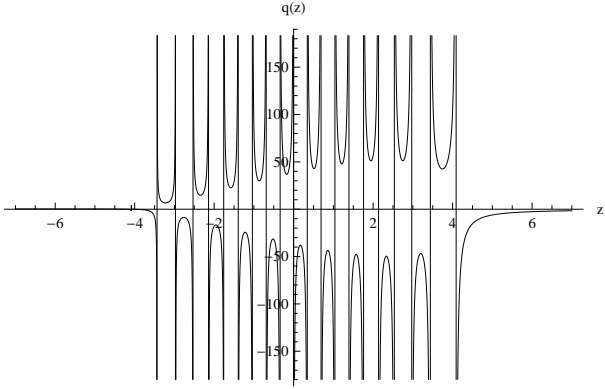}
\caption{The deceleration parameter $q(z)$ (\ref{31}) \textit{versus} the deviation $z$ for the potential energy (\ref{40}) with $n = 10$.}
\end{figure*} 

\subsection{The universe of type II}
Let us consider the quantum universe described by the wavefunction $| f \rangle_{2}$ (universe of type II). The solution 
$| f \rangle_{2}$ from (\ref{36}) as a function of $z = a - M$, where $a$ is a real variable, is shown in Fig.~3 for $n = 10$.
Its real $Re\, | f \rangle_{2}$ and imaginary $Im\, | f \rangle_{2}$ parts oscillate in the interval $|z| < M$ and are
shifted in the phase with respect to each other by $\frac{\pi}{2}$. The function $Re\, | f \rangle_{2}$ decreases exponentially outside this interval,
while $Im\, | f \rangle_{2}$ diverges exponentially at $|z| \rightarrow + \infty$. In the interval bounded by the values $|z| \leq M$,
the function $| f \rangle_{2}$ can be normalized. The normalization constant will depend on quantum number $n$.

\begin{figure*}
\includegraphics[width=8cm]{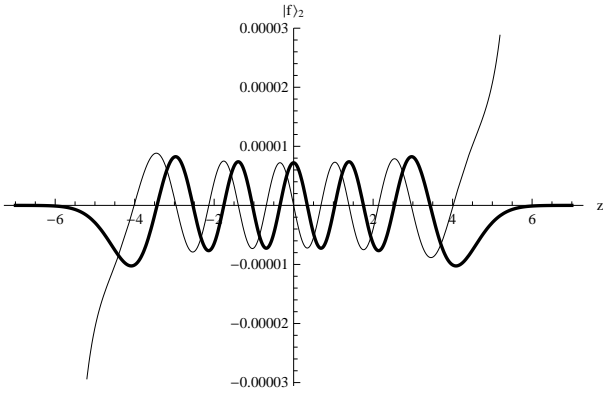}
\caption{The real (boldface curve) and imaginary (thin curve) parts of the function $| f \rangle_{2}$ from (\ref{36}) \textit{versus} the deviation $z$ for $n = 10$.}
\end{figure*} 

Using Eq. (\ref{38}), we obtain the expression for $Q$ (\ref{18}) as a function of $i z$, where the deviation $z$ is a real variable as before,
\begin{equation}\label{45}
Q(i z) = - (2 n + 1) + 2 (n + 1) \frac{H_{-n-2}(i z) H_{-n}(i z)}{H_{-n-1}^{2}(i z)}.
\end{equation}
The potential well (\ref{17}) takes the form
\begin{equation}\label{46}
U(z) = \frac{1}{2} z^{2} - (n + 1) \frac{H_{-n-2}(i z) H_{-n}(i z)}{H_{-n-1}^{2}(i z)}.
\end{equation}
It is a complex function of the form 
\begin{equation}\label{46a}
U(z) = U_{R}(z) + i U_{I}(z)
\end{equation}
where $U_{R}(z)$ and $U_{I}(z)$ are real functions.

The evolution of the universe with complex potential well (\ref{46a}) can be described in terms of the formalism with complex scale factor
\begin{equation}\label{46b}
a = a_{R} + i\, a_{I},
\end{equation}
where $a_{R}$ and $a_{I}$ are real functions of time $T$. The possibility of an introduction of a complex metric tensor and its relation
to real physical gravitational field was studied, e.g., in Refs. \cite{Mof,Cha} (see also references therein). Taking the common point of view,
we shall assume that the physical gravitational field is described by the real part of the metric (\ref{1}) (the real line element).
In our model, the necessity to pass to complex variable $a$ is connected to the complexity of the wavefunction $|f\rangle_{2}$.
Since the real and imaginary parts of this function vanish at different points (see Fig.~3), the real physical quantities, such as
the kinetic energy, the potential well, and the deceleration parameter appear to be free of discontinuities which are typical 
for the real function $|f\rangle_{1}$ in the region $-M < z < M$ (see Figs.~1 and 2).

The energy conservation law (\ref{16}) can be rewritten in the form of two conditions
\begin{equation}\label{26}
\frac{1}{2} \left[ \left(\frac{da_{R}}{dT}\right)^{2} - \left(\frac{da_{I}}{dT}\right)^{2} \right] + U_{R} = 0, \quad
\frac{da_{R}}{dT}\, \frac{da_{I}}{dT} + U_{I} = 0.
\end{equation}
Hence it follows that there are two solutions for the parts of kinetic energy related to the change of $a_{R}$ and $a_{I}$ with time $T$,
\begin{equation}\label{27}
\left(\frac{da_{R}}{dT}\right)_{\pm}^{2} = - U_{R} \pm \sqrt{U_{R}^{2} + U_{I}^{2}}, \quad
\left(\frac{da_{I}}{dT}\right)_{\pm}^{2} = \frac{U_{I}^{2}}{-U_{R} \pm \sqrt{U_{R}^{2} + U_{I}^{2}}}.
\end{equation}

The potential well (\ref{46a}) depends only on the real part of the scale factor.
Therefore instead of $z = a - M$ one must take $z = a_{R} - M$ in Eq. (\ref{46}). From Eqs. (\ref{26}) and (\ref{27}), it follows that
the imaginary part $a_{I}$ must be expressed in terms of the real part $a_{R}$,
\begin{equation}\label{47}
a_{I} = \int_{0}^{a_{R}} \! dx \frac{U_{I}(x)}{U_{R}(x) \mp \sqrt{U_{R}^{2}(x) + U_{I}^{2}(x)}},
\end{equation}
with the boundary condition $a_{I}(a_{R} = 0) = 0$. Thus, in such a model,
all elements of the complex spacetime are expressed via one real parameter $a_{R} = a_{R}(T)$.

The complexity of the spacetime metric leads to interference of the kinetic energies 
$K_{R,I}^{\pm} \equiv \frac{1}{2} \left(\frac{d a_{R,I}}{d T} \right)_{\pm}^{2}$ described by Eqs. (\ref{27}). This interference smoothes out
behaviour of these energies near the points $|z| = z_{0}$, where $U_{R} = 0$. In this case, the motion is always realized in real time $T$,
since the domain with the Euclidean signature is found to be inaccessible.

The real and imaginary parts of the energy (\ref{46}) as functions of $z$ are plotted in Fig.~4 for the quantum number $n = 10$.
The general behaviour of $U_{R}$ and $U_{I}$ with respect to $z$ is the same for arbitrary values of the quantum number
$n$, from $n \sim 1$ up to $n \gg 1$. The points, where $U_{R}$ and $U_{I}$ have extrema or vanish, are determined by $n$ and $M$.
So we can conclude that in the interval $|z| < M$, the energy $U_{R}(z)$ is well approximated by the expression: 
$U_{R} = \frac{1}{2} z^{2} - \left(n + \frac{1}{2} \right)$. It vanishes at the points $z_{0} \approx \pm \sqrt{2 (n + 1)}$.
The imaginary part $U_{I}$ has extrema at these points (cf. the wavefunction $| f \rangle_{2}$ in Fig.~3). It vanishes at the points
$z = 0$ and $|z| = + \infty$. The real part of the potential energy is negative, $U_{R} < 0$, in the region $|z| < z_{0}$ and positive, 
$U_{R} > 0$, for $|z| > z_{0}$. We have $U_{R} \rightarrow + \infty$ at $|z| \rightarrow + \infty$.
The imaginary part of the potential energy is positive, $U_{I} > 0$, at $z < 0$ and negative, $U_{I} < 0$, on the semiaxis $z > 0$.
The inequality $|U_{R}| \gg |U_{I}|$ holds in the whole range of $z$, except the points near $|z| = z_{0}$.

\begin{figure*}
\includegraphics[width=8cm]{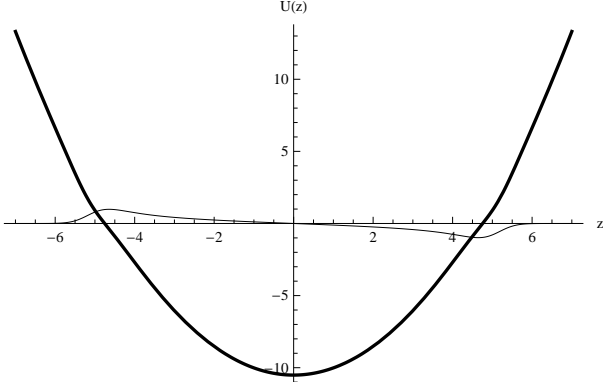}
\caption{The real (boldface curve) and imaginary (thin curve) parts of the potential well $U(z)$ (\ref{46}) \textit{versus} the deviation 
$z = a_{R} - M$ for $n = 10$.}
\end{figure*} 

In Fig.~5, the real $K_{R}^{+}$ and imaginary $K_{I}^{+}$ parts of the kinetic energy (\ref{27}) of type $(+)$ for $n = 10$ are shown.
In the whole range of the deviation $z = a_{R} - M$, both these energies are positive and describe the motion in real time.
In the points $|z| = z_{0}$, they equal each other in accordance with Eqs. (\ref{26}). In the point $z = 0$, the energy $K_{R}^{+}$  
has a maximum equal to $\left(n + \frac{1}{2} \right)$, while the energy $K_{I}^{+}$ vanishes.
In the interval $|z| < z_{0}$, the condition $K_{R}^{+} \gg K_{I}^{+}$ is satisfied. In the domain $|z| > z_{0}$, where $U_{R} > 0$,
the energy $K_{R}^{+} \rightarrow 0$ for $|z| \rightarrow + \infty$, while the energy 
$K_{I}^{+} \rightarrow U_{R}$.

\begin{figure*}
\includegraphics[width=8cm]{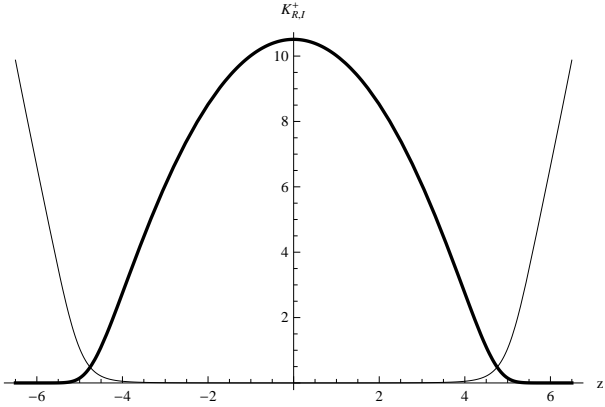}
\caption{The real $K_{R}^{+}$ (boldface curve) and imaginary $K_{I}^{+}$ (thin curve) parts of the kinetic energy (\ref{27}) \textit{versus} the 
deviation $z = a_{R} - M$ for $n = 10$.}
\end{figure*} 

The kinetic energy $K_{R}^{+}$ increases in the interval $- \infty < z < 0$. It means that the internal forces perform the positive work on the
universe accelerating the expansion. This work is analogous to the work of the forces of dark energy. On the contrary,
the kinetic energy $K_{I}^{+}$ decreases in this interval, the negative work is done on the universe. As a result,
the expansion decelerates. This is equivalent to the work of the attractive forces of dark matter, that is, matter-energy that
does not contribute into the mass $M$.

The energy $K_{R}^{+}$ decreases in the interval $0 < z < + \infty$, demonstrating that the work performed on the universe is negative. 
In this way,
the presence of the additional source of gravitational attraction which is not associated with the mass $M$ is imitated.
This additional source can be again identified with dark matter. Inversely, the energy $K_{I}^{+}$ increases in this interval,
the performed work is positive, as under the action of the forces of dark energy.

So, the kinetic energies $K_{R}^{+}$ and $K_{I}^{+}$ show what work should the universe do in order to overcome the action of
the internal forces of repulsion (dark energy) and attraction (dark matter) which exist simultaneously and compete with each other
at all stages of the evolution of the universe. Which of these forces reveals itself as dark energy (dark matter) depends on the sign
of the corresponding force. Whether the expansion of the universe is accelerating or decelerating depends on the relation between
the forces performing the work, causing acceleration or deceleration.

The plots of the real $K_{R}^{-}$ and imaginary $K_{I}^{-}$ parts of the kinetic energy (\ref{27}) of type $(-)$ would be 
the mirror images of the plots in Fig.~5 with regard to substitutions $K_{R}^{+} \rightarrow - K_{I}^{-}$ and
$K_{I}^{+} \rightarrow - K_{R}^{-}$. Both energies are negative in the whole range of the deviation $z$ and
the motion can be described in imaginary time $\xi = -i T$.
The analysis of the solution of type $(+)$ given above remains valid for the solution of type $(-)$ after a formal substitution
$T \rightarrow \xi$.
It means that gravitational and antigravitational forces which perform work on the universe analogous to dark matter and dark energy
can exist in the spacetime with the Euclidean-signature metric as well.

Thus, the presence of the imaginary part $U_{I}$ in the potential well (\ref{46a}) and in Eqs. (\ref{26}) indicates that the processes 
of absorption and release of energy pumping over between the states with an effective attraction and repulsion of matter are taking place
in the system.

In Fig.~6, the real $q_{R}$ and imaginary $q_{I}$ parts of the deceleration parameter (\ref{31}) are shown as the functions of the deviation $z$
for the potential well (\ref{46}) with $n = 10$.
In the region $|z| \leq M$, where $|q_{R}| \gg |q_{I}|$ (i.e. $|q_{I} / q_{R}|_{z = 0} \approx 0.02$), the contribution from $q_{I}$ can be 
neglected. In this stage, the universe expands with deceleration, since the antigravitational action of the forces performing the
positive work is not enough to overcome attraction of ordinary and dark matter. The value $q_{R} (z = 0) = 1$ reproduces 
the results of general relativity \cite{Wei72}. At the point $z = 0$, we have $a_{R} = M$. In the region  $a_{R} \approx 2 M$, the
redistribution of energy takes place in the universe as demonstrated by the peaks on the curves $q_{R}$ and $q_{I}$ in Fig.~6.
The forces of attraction and repulsion compete with each other at $a_{R} < 2 M$, where $q_{R} > 0$ and $q_{I} < 0$.
At reaching the region $z > M$, where $a_{R} > 2 M$, both parts of the deceleration parameter become negative, demonstrating
that the expansion of the universe is accelerating.
Starting from the point $z \simeq 1.5 M$ ($z = 6$ for $n = 10$), the parameter $q_{I}$ vanishes and the rate of expansion is 
described only by the real part $q_{R} < 0$. In the limit $z \rightarrow + \infty$,
the forces of attraction and repulsion will exactly compensate each other.

\begin{figure*}
\includegraphics[width=8cm]{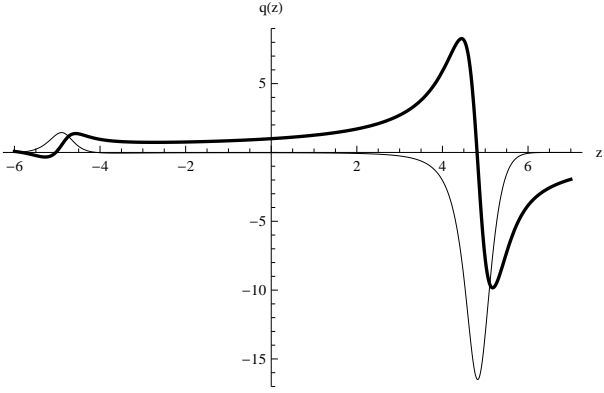}
\caption{The real $q_{R}$ (boldface curve) and imaginary $q_{I}$ (thin curve) parts of the deceleration parameter (\ref{31}) \textit{versus} the deviation $z = a_{R} - M$
for the potential energy (\ref{46}) with $n = 10$.}
\end{figure*} 

Again, as in the case of the complex metric tensor, one can accept that only the real part of the deceleration parameter $q_{R}$ 
is a physically measurable quantity. The imaginary part $q_{I}$ plays a role of regularizing factor which allows to exclude discontinuities
caused by vanishing of the real part of the function $| f \rangle_{2}$ in isolated points.

Let us compare the predictions of the quantum model under consideration with the observations in our universe. In the modern era,
the scale factor (radius of the universe) $a_{R} \sim 10^{28}$ cm and the mass $M \sim 10^{80}$ GeV of matter 
in the observed part of our universe are estimated up to coefficient less than $O(10)$. 
Such a radius $a_{R}$ roughly coincides with the Hubble radius, while the mass $M$ is estimated by the quantity of matter
with the critical density $\rho_{c} \approx 10^{-5}$ GeV/cm$^{3}$ contained in the Hubble volume $\approx 2 \pi^{2} a_{R}^{3}$.
In dimensionless units, which we use in this paper, these parameters prove to be of the same order of 
magnitude, $a_{R} \sim 10^{61}$ and $M \sim 10^{61}$. According to observations and theoretical estimations,
the transition from the matter dominated phase to the dark energy dominated universe took place at the redshift $\approx 0.6$.
It does not contradict with the obtained condition $2 < a_{R}/M < 10$ (see Fig.~6) which determines the stage of transition to the phase of 
accelerating expansion of the universe.

\section{Conclusion}
In this paper, the evolution of the universe is studied in exactly solvable dynamical quantum model with the Robertson-Walker metric.
It is shown that the equation of motion which describes the expansion or contraction of the universe can be represented in the form
of the zero total energy conservation law (\ref{16}) for the particle being an analogue of the universe. The analogue particle has an arbitrary
mass and moves in the potential well (\ref{17}) under the action of the internal force (\ref{23}) which takes into account the curvature of space,
mass term, and gradients (pressures) of classical and quantum gravitational sources. The quantum source (\ref{18})
emerges as a result of the evolution of the phase $S_{k} (a)$ of the state vector (\ref{15}), which describes the geometrical properties
of the quantum universe, in the space of the scale factor. Equation (\ref{22}) for the phase $S_{k} (a)$ is non-linear, and contains the
information about the curvature of space and quantum states of matter in the universe.

In a particular case of matter in the form of dust, this non-linear equation has the analytical solutions of two types: (i) real solution
for the Euclidean phase $S_{E} = -i S_{1}$
(\ref{37}) which corresponds to the real state vector $| f \rangle_{1}$ from (\ref{36}) (universe of type I); (ii) complex
solution $S_{2}$ (\ref{38}) for the state vector $| f \rangle_{2}$ from (\ref{36}) in the space of complex scale factor (universe of type II).

It is shown that, as the universe of type I evolves,
it subsequently passes through the phases (stages) of the accelerating and decelerating expansion. The phases change each other
with the interval of the order of Planck length in displacement scale of $z = a - M$. 

The description of the motion of the analogue particle as a mathematical equivalent of the evolving universe of type II
is characterized by two types of possible solutions for the real and imaginary parts of the kinetic energy (\ref{27}) of types $(+)$ and
$(-)$. The solution of type $(+)$ describes the motion of the analogue particle in real time $T$, while the solution of type $(-)$
corresponds to imaginary time $\xi = - i\,T$. The changes of the real and imaginary parts of the kinetic  
energy of one type during the evolution of the universe demonstrate that
the internal forces simultaneously perform both the positive work on the universe (e.g., the energy $K_{R}^{+}$ increases as in Fig.~5),
which is analogous to the work of the forces of dark energy, and the negative work (the energy $K_{I}^{+}$ decreases),
which is similar to the work of the attractive forces of dark matter. The general character of the expansion of the universe of type II
at a definite instant of time (parametrized by the deviation $z = a_{R} - M$ in Fig.~6) depends on which of the works dominates.
As in the case of the universe of type I (see Fig.~2), the expansion of the universe becomes accelerating after reaching the region
$a_{R} > 2 M$. This result does not contradict the data on the expansion of our universe in the modern era and predicts that the forces of 
attraction and repulsion will exactly compensate each other in infinite future ($a \rightarrow + \infty$). 

In approach under consideration, the change of the regimes of
the expansion of the universe of both types is not caused by the action of some specific material carrier of mass-energy. 
This phenomenon reflects a quantum nature of the universe. The equations of the quantum theory (\ref{16}) and  (\ref{16a})
transform into the Einstein-Friedmann equations of general relativity (\ref{23a}) without dark energy in the limit $Q_{k}(a) \rightarrow 0$.

Thus, it appears that the quantum universe is such that, during its expansion, it decelerates, then accelerates, or vice versa, spontaneously.
The cause of the expansion and change of its regimes is a special form of the potential well
(\ref{17}) in which the universe as a whole is moving.

\end{document}